\documentclass[aps,prl,twocolumn,superscriptaddress]{revtex4}

\usepackage{graphicx,amsmath}

\begin{document}

\title{Many-Body Electrostatic Forces Between Colloidal Particles at Vanishing Ionic Strength}
\author{Jason W. Merrill}
\affiliation{Yale University, Department of Physics}
\email[]{jason.merrill@yale.edu}
\author{Sunil K. Sainis}
\affiliation{The Rowland Institute}
\author{Eric R. Dufresne}
\email[]{eric.dufresne@yale.edu}
\affiliation{Yale University, Departments of Mechanical Engineering, Chemical Engineering, and Physics}

\date{\today}

\begin{abstract}
	Electrostatic forces between small groups of colloidal particles are measured using blinking optical tweezers. When the electrostatic screening length is longer than the inter-particle separation, forces are found to be non-pairwise additive. Both pair and multi-particle forces are well described by the linearized Poisson-Boltzmann equation with constant potential boundary conditions. These findings may play an important role in understanding the structure and stability of a wide variety of systems, from micron-sized particles in oil to aqueous nanocolloids.
\end{abstract}


\maketitle

Colloidal suspensions offer a unique opportunity to observe the relationship between microscopic interactions and macroscopic thermodynamic behavior because the component particles are both large enough to observe directly with light microscopy and small enough to come to thermodynamic equilibrium over experimental timescales. Colloids display a rich range of phase behavior including equilibrium gases, liquids, and crystals and nonequilibrium glasses and gels. Controlling and predicting this phase behavior has applications from ensuring the stability of paint to directing the self assembly of novel optical materials.

Predicting phase behavior requires an understanding of the interactions of the component particles. Forces between pairs of charged surfaces have been extensively characterized using direct force measurements \cite{israelachvili-1992}. Alternatively, observations of bulk correlations can be used to extract an ``effective pair potential,'' but this pair potential is found to vary with the density of the suspension \cite{Brunner:2002p2239,RojasOchoa:2008p2215}. This suggests that charged colloidal particles do not, in fact, interact only through a pair potential. Many-body interactions evidently play an important role in determining macroscopic behavior.

To bridge the gap between pair interactions and bulk behavior, we explore the regime of few-body interactions by using blinking optical tweezers to directly measure electrostatic forces between sets of several isolated charged colloidal particles at low ionic strength. For screening lengths much shorter than the inter-particle separation, the measured forces are nearly pairwise additive; however, for screening lengths longer than the inter-particle separation, there are significant deviations from pairwise additivity.

Our system consists of radius $a =$ 600 nm PMMA spheres suspended in hexadecane, a non-polar solvent. The spheres are coated in PHSA in order to stabilize them against aggregation \cite{Antl:1986p2263}. Both the particle charge and screening length are adjusted by adding variable concentrations of a reverse micelle forming surfactant, NaAOT, as described elsewhere \cite{Eicke:1989p2268,Hsu:2005p2267,Sainis:2008p13334,Roberts:2009p2242}.

Forces are extracted from the trajectories of freely interacting particles, which are repeatedly trapped and released using blinking optical tweezers \cite{Crocker:1994p2119} created with a 1064 nm laser. The particles are more than 30 particle diameters from the nearest wall. A detailed explanation of the apparatus is given in \cite{Sainis:2008p2262}.  The physical ideas underlying the technique are developed in \cite{Sainis:2007p2048}. The central result of that paper is a relation between the force, $f$, and the two statistical parameters that describe ensembles of short-time Brownian trajectories, the drift velocity, $v_d$, and the diffusion coefficient, $D$:
\begin{equation}
	f = k_B T D^{-1} v_d.
	\label{eq:force-measurement}
\end{equation}
This relation holds for a single particle in one dimension, or separately for each of the hydrodynamic normal modes of a multi-particle system in several dimensions. The range of validity of this expression is explored in \cite{Sainis:2007p2048}.

\begin{figure*}[htb]
\includegraphics[width=.9\textwidth]{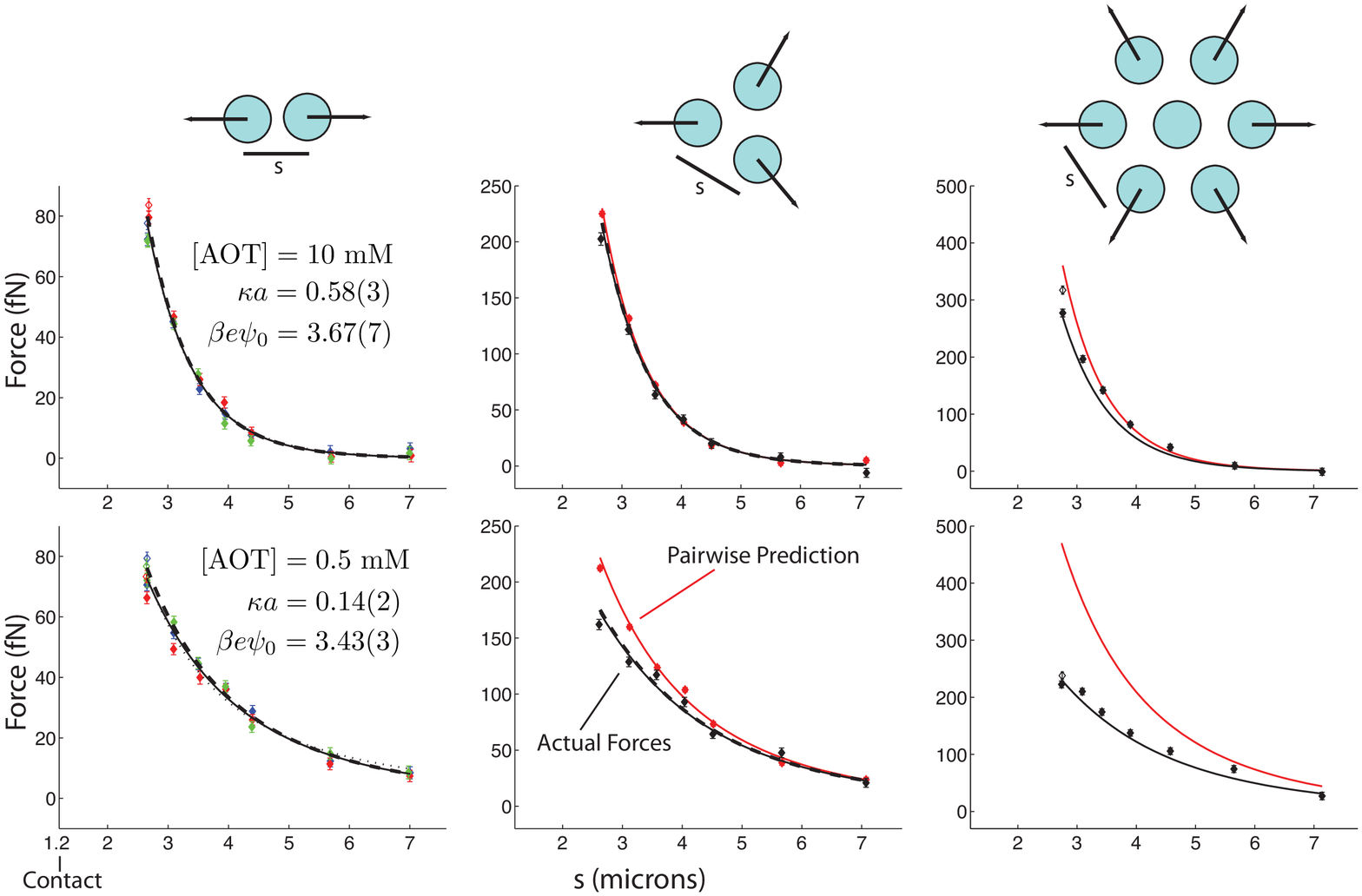}
\caption{\label{fig:forces} \emph{Direct Measurement of Non-Pairwise Electrostatic Interactions.} Forces on beads in pair (first column), equilateral (second column), and hexagonal (third column) configurations, at AOT concentrations of 10 mM (first row) and 0.5 mM (second row). 
The arrows on the particle configurations in the first row indicate the form of the breathing modes used for analysis.  Breathing modes are normalized so that sum of the squares of the particle displacements is unity.
For the pair measurements, different colored points represent different pairs in the same sample.
The dotted, solid, and dashed lines are fits to constant charge density, constant potential, and a simple approximation of constant potential based on Eq.\ \ref{eq:potential-charge-bulk}.
For the equilateral and hexagonal configurations, black points are measured forces on the breathing mode and red points are a direct pairwise sum of the measured pair forces.
The red lines are pairwise sums of the constant potential pair fits. Constant potential predictions for the force on the breathing mode based on fits to the pair data are shown as black lines.
The solid line is based on the full numerical solution while the dashed line is based on Eq.\ \ref{eq:potential-charge-bulk}.}
\end{figure*}

We measure forces between beads in pair, equilateral triangle, and hexagonal configurations, as depicted in Fig.\ \ref{fig:forces}. The first column of Fig.\ \ref{fig:forces} shows the measured force as a function of separation, $s$, between pairs of beads at two different surfactant concentrations: $[\mathrm{AOT}] =$ 10 mM (top) and 0.5 mM (bottom). At each separation, three beads are chosen for pair measurements. The forces between each of the three possible pairs formed from this set are plotted with different colors. For a given concentration and separation, forces between each of the pairs are nearly identical, which suggests that each of the beads has nearly the same charge. The force at the smallest measured separation is roughly the same for both concentrations, but the interaction is screened much more rapidly at the higher surfactant concentration.

Next, the same three beads are formed into an equilateral triangle. The forces on the breathing mode are calculated using Eq.\ \ref{eq:force-measurement} and displayed in the second column of Fig.\ \ref{fig:forces}. The previously measured pair forces are used to make a pairwise prediction of the force on the equilateral triangle. Since the same beads are measured at the same separations for both the pair and equilateral geometry, this experiment provides a strong, model-independent test of pairwise additivity. At 10 mM, agreement with the pairwise prediction is very good, whereas at 0.5 mM, there are significant deviations from pairwise additivity.

Finally, the same three beads along with four others are formed into a hexagon with a particle at its center. The forces on the breathing mode are displayed in the third column of Fig.\ \ref{fig:forces}. There is a striking disagreement with the pairwise additive prediction at 0.5 mM, and a small deviation at 10 mM.

At the end of the experiment, after measurements in all three configurations are complete, force measurements at the smallest separation in the pair and hexagonal configurations are repeated. These measurements are plotted as open symbols. Their agreement with measurements from the beginning of the experiment suggests that the forces do not drift noticeably over time.

These forces are electrostatic. In the presence of mobile ions, the electrostatic potential, $\psi$, obeys the Poisson-Boltzmann equation (PBE):
\begin{equation}
	\beta e \nabla^2 \psi = \kappa^2 \sinh(\beta e \psi),
\label{eq:pbe}\end{equation}
where $\kappa^2 = 2 n_\infty \beta e^2/\epsilon$ is the ionic strength, which is proportional to the density of ions in the bulk, $n_\infty$. The quantity $\kappa^{-1}$, called the screening length, sets the range of electrostatic interactions. Since the screening length scales as $\kappa^{-1} \sim n_\infty^{-1/2}$, the dimensionless parameter $\kappa a$ can be the same for micron-sized particles in oil and nanocolloids in water.

As a result of the non-linearity of the PBE, its solutions cannot be superposed. Thus, forces are not generally pairwise additive. Russ \textit{et al.}\ numerically solved the PBE with constant charge density boundary conditions for two and three spheres to find a three-body potential \cite{Russ:2002p2228}. This theory was used to explain deviations from pairwise additivity found in experiments conducted in water with $\kappa a \approx 2$ \cite{Brunner:2004p2134}. The constant charge density PBE fails to describe our data, where $\kappa a < 1$, because it predicts deviations from pairwise additivity that increase with ionic strength: the opposite of the trend observed in our system.

\begin{figure}[hbt]
\includegraphics[width=.9\columnwidth]{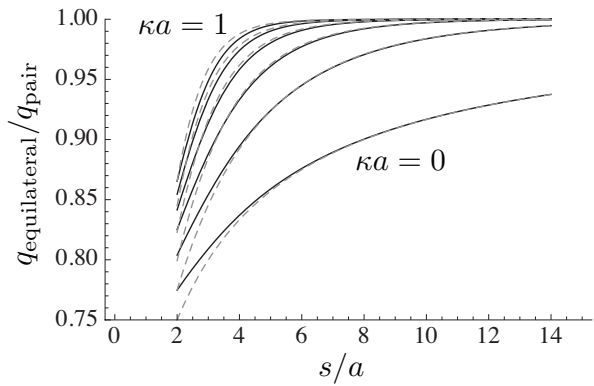}
\caption{\label{fig:isolated-charge} \emph{Charge reduction in the presence of a third particle} Ratio of the charge per particle on a set of three constant potential spheres arranged in an equilateral triangle of side length $s$ to the charge per particle on a pair of similar spheres separated by length $s$ for six values of the ionic strength: $\kappa a =$ 0, 0.2, 0.4, 0.6, 0.8 and 1.0. The solid line represents a numerical solution of the LPBE, and the dashed line represents an analytical approximation based on a single point charge at the center of each sphere, Eqs.\ \ref{eq:potential-charge-2}, \ref{eq:potential-charge-3}.}
\end{figure}

Non-linear governing equations are not necessary for non-pairwise additive forces. Indeed, forces between finite size conductors held at constant potential are non-pairwise additive even in free-space electrostatics, which is a perfectly linear theory. In fact, we find that the \textit{linearized} Poisson-Boltzmann equation (LPBE) is sufficient for describing all of our observations including deviations from pairwise additivity.

In order to maintain constant potential, the total amount of charge on a pair of spheres must change as they are brought together. As a result, the force law between a pair of conducting spheres is different from the force law between point charges \cite{Soules:1990p2110}. Similarly, as a third sphere approaches a pair, the charge will change even further. This is the reason for non-pairwise additivity of forces. This is illustrated in Fig.\ \ref{fig:isolated-charge}, where we plot the ratio of the charge on each sphere in an equilateral triangle of side length $s$ to the charge on each sphere in a pair of the same separation. The charge is calculated for several values of $\kappa a$ by numerically solving the LPBE using a simple singularity method \cite{Phillips:1995p2199}.

Fits to the constant potential pair force law \cite{Phillips:1995p2199} are shown as solid lines in Fig.\ \ref{fig:forces}. The resulting values of $\kappa a$ and $\psi_0$ are shown in Fig.\ \ref{fig:forces}.  For comparison, fits to the constant charge density force law are shown as dotted lines. Clearly, it is impossible to distinguish between the two force laws based only on these pair force data. However, the two fits give different values for the screening length and the charge or surface potential. This leaves an essential ambiguity about the underlying physics.

The many body data resolve this ambiguity. Based on the surface potential and screening length extracted from the constant potential fits, we calculate the expected forces on the breathing mode of the equilateral and hexagonal configurations, which are plotted as solid black lines in Fig.\ \ref{fig:forces}. This linear constant potential model gives excellent quantitative predictions for the deviations from pairwise additivity observed at $[\mathrm{AOT}] = 0.5$ mM. The constant charge density model predicts no deviations from pairwise additivity. At 10 mM, where both predicted and observed deviations from pairwise additivity are small, it is not clear that the constant potential model is much better than simply assuming pairwise additivity. We emphasize that all the information used to calculate the expected forces in the equilateral and hexagonal configurations comes from fits to the pair data---no additional free parameters are introduced.

In order to understand why this constant potential model predicts larger deviations from pairwise additivity at lower ionic strength, it is helpful to consider a simple analytical approximation. Model each particle with a single point charge at its center with a value that is free to change in order to make the particle surfaces roughly constant potential. Solving the LPBE shows that the potential on the surface of a sphere due to a point charge $q$ at its center is $q e^{-\kappa a}/4 \pi \epsilon a$. The potential due to a point charge outside the sphere, a distance $s$ from its center, varies over the surface of the sphere, but can be roughly approximated by the value of the potential at its center, $q e^{-\kappa s}/4 \pi \epsilon s$, so long as $s \gg a$. The total potential on the surface of particle $i$ is thus approximately
\begin{equation}
	\psi_{0,i} = \frac{1}{4 \pi \epsilon} \left(
		q_i \frac{e^{-\kappa a}}{a} + \sum_{j \ne i} q_j \frac{e^{-\kappa s_{ij}}}{s_{ij}  } \right),
	\label{eq:potential-charge-bulk}
\end{equation}
where $s_{ij}$ is the separation between particles $i$ and $j$. If we demand that this potential takes on some fixed value, then this is a system of linear equations that can easily be solved for the charge at the center of each particle as a function of the configuration, $s_{ij}$. The solutions for the charge per particle in pair and equilateral configurations are
\begin{eqnarray}
	\label{eq:potential-charge-2}
	q_\mathrm{pair} & = & 4 \pi \epsilon \psi_0 (e^{-\kappa a}/a + e^{-\kappa s}/s)^{-1} \\
	\label{eq:potential-charge-3}
	q_\mathrm{equilateral} & = & 4 \pi \epsilon \psi_0 (e^{-\kappa a}/a + 2 e^{-\kappa s}/s)^{-1}
\end{eqnarray}
It is important to note that the singular charge, $q$, is only equal to the actual charge on a particle, $Q$, in the Coulomb limit. In general, for isolated particles they are related by $Q = q e^{-\kappa a} (1 + \kappa a)$, a result that can be obtained from Gauss's law. Predictions based on this approximation for the total charge as a function of separation are plotted as dashed lines in Fig.\ \ref{fig:isolated-charge}. The approximation is apparently quite good.

In the equilateral configuration, the force on any one of the spheres should scale as $q^2$, so the ratio of the true force to the pairwise force is, in this approximation,
\begin{equation}
	\frac{f_\mathrm{equilateral}}{f_\mathrm{pairwise}} = \left(\frac{q_\mathrm{equilateral}}{q_\mathrm{pairwise}} \right)^2 = \left(\frac{1 + e^{-\kappa (s - a) }a/s}{1 + 2 e^{-\kappa (s - a) }a/s}\right)^2.
	\label{eq:force-ratio}
\end{equation}
This shows how the non-pairwise additivity depends on the three relevant length scales: $\kappa^{-1}$, $s$, and $a$. The expected forces based on this approximation are plotted as dashed lines in the second column of Fig.\ \ref{fig:forces}, and are in excellent agreement with both the numerical calculation and the data over the full range of measured separations. This analytical approximation does not take polarization into account, whereas the numerical calculation does. Their agreement demonstrates that, in these experiments, polarization is not important compared to the change in total charge.

Assuming the surface potential is independent of configuration leads to an accurate prediction of the observed deviations from pairwise additivity. Assuming constant charge density is evidently wrong, at least for the measurements at lower ionic strength. But why should the surface potential be constant? After all, these are isolated dielectric particles, not conductors hooked up to a battery.

In fact, the idea of constant potential colloid surfaces is far from new. In their foundational 1948 work on colloidal interactions, Verwey and Overbeek argued that colloidal particles should be at constant surface potential whenever surface ions are in equilibrium with bulk ions, and criticized earlier assumptions of constant charge as artificial \cite{verwey-1948}. Finding equilibrium by equating the bulk and surface ion chemical potentials led them to the Nernst equation, $\beta e \psi_0 = \ln (n_\infty/n_\mathrm{ref})$, where $n_\mathrm{ref}$ is a material parameter. This suggests that the surface potential depends only on material properties and the bulk concentration of ions, and not on geometry or any interactions between particles. However, they explicitly ignore the surface configurational entropy, which can couple the surface potential to the fields to push the system toward constant charge \cite{Ninham:1971p2203,Chan:1983p2232,Behrens:1999p2220}. Different models of the surface chemistry make different predictions for the surface configurational entropy, so the agreement of constant potential boundary conditions with our data places constraints on the surface chemistry. In particular, it appears that a surface with both positive and negative dissociable ions---an amphoteric surface---should be in better accord with the Nernst equation and thus closer to constant potential than a surface with only a single dissociable ion species \cite{Larson:2000p2235}.

With few exceptions \cite{Verhoeff:2007p2266}, previous predictions of colloidal structure have uniformly assumed constant charge density boundary conditions \cite{Hone:1983p2212,Robbins:1988p2260,Dobnikar:2003p2136,Leunissen:2005p2259,RojasOchoa:2008p2215}. Our experimental evidence for constant potential boundary conditions demands a fresh look at the bulk structure of charged colloidal suspensions. As a first step, we have calculated the charge per particle in a bulk system of identical constant potential particles as a function of separation using the PB cell model \cite{Alexander:1984p2211,Trizac:2003p2213}. At very low ionic strength, the charge per particle drops significantly even at modest volume fractions. For $\kappa a = 0.14$ and $\beta e \psi_0 = 3.4$, the measured values at $[\mathrm{AOT}] = 0.5$ mM, the charge per particle at 10\% volume fraction is reduced by a factor of 6 compared to an isolated particle. Thus, we expect constant potential boundary conditions to have important implications for predictions of aggregation and self assembly in a wide range of systems where the screening length is larger than the particle size. An interesting case to consider is binary oppositely charged colloids at $\kappa a \approx 1$, where the self assembly of diamond-like structures has been observed \cite{Kalsin:2006p2255}. In that case, constant potential boundary conditions should have the opposite effect on particle charge---as the volume fraction increases we expect the charge per particle to increase. Much theoretical work is needed.

We acknowledge support from the National Institute for Nano Engineering at Sandia National Laboratory and an NSF CAREER grant to E.R.D. (CBET-0547294).

\end{document}